\documentstyle[12pt,epsfig]{article}

 \large\normalsize

\newcommand{\be}{\begin{equation}}
\newcommand{\ee}{\end{equation}}
\newcommand{\tr}{\mathop{\rm Tr}\nolimits}
\def\spose#1{\hbox to 0pt{#1\hss}}
\def\ltapprox{\mathrel{\spose{\lower 3pt\hbox{$\mathchar"218$}}
 \raise 2.0pt\hbox{$\mathchar"13C$}}}
\def\gtapprox{\mathrel{\spose{\lower 3pt\hbox{$\mathchar"218$}}
 \raise 2.0pt\hbox{$\mathchar"13E$}}}
\def\reff#1{(\ref{#1})}
\def\bvx{\vec{\mbox{$x$}}}

\newcommand{\1}{1\!\!\!\bot}
\newcommand{\ba}{\begin{eqnarray}}
\newcommand{\ea}{\end{eqnarray}}

\begin{document}

\title{Gribov Copies in the Minimal Landau Gauge: \\
         the Influence on Gluon and Ghost Propagators}
 
\author{{\small Attilio Cucchieri}\thanks{~Present address:
Gruppo APE -- Dipartimento di Fisica,
Universit\`a di Roma ``Tor Vergata'',
Via della Ricerca Scientifica 1, I-00133 Roma, ITALY
(e-mail: {\tt cucchieri@roma2.infn.it}).} \\ [-0.1cm]
  {\small ZiF--Zentrum f\"ur Interdisziplin\"are Forschung
          der Universit\"at Bielefeld} \\ [-0.1cm]
  {\small Wellenberg 1, D-33615 Bielefeld, GERMANY} \\ [-0.1cm]
  {\small and} \\ [-0.1cm]
  {\small Fakult\"at f\"ur Physik, Universit\"at Bielefeld} \\ [-0.1cm]
  {\small Universit\"atsstrasse, D-33615 Bielefeld, GERMANY} }
 
\date{August 4, 1997}

\maketitle

\begin{abstract}

We study the influence of Gribov copies on gluon and ghost propagators,
evaluated numerically in pure $SU(2)$ lattice gauge theory
in the minimal Landau gauge.
Simulations are done at four different values of
$\beta$ (namely $\beta = 0\mbox{,}\, 0.8\mbox{,}\,
1.6\,$ and $\, 2.7\,$) and for volumes up to $16^4$
(up to $24^4$ at $\beta = 1.6$).
For the gluon propagator, Gribov noise seems to be of the
order of magnitude of the numerical accuracy, even
at very small values of the coupling $\beta$.
On the contrary, for the ghost propagator, Gribov noise is 
clearly observable for the three values of $\beta$
in the strong-coupling regime. In particular, data
corresponding to the minimal Landau gauge are always
smaller than those obtained in a generic Landau gauge.
This result can be qualitatively explained.

\end{abstract}

 
Gauge theories, being invariant under local gauge transformations,
are systems with redundant dynamical variables, which
do not represent true dynamical degrees of freedom.
The objects of interest are not the gauge fields
themselves, but rather the classes (orbits) of gauge-related fields.
The elimination of such redundant gauge degrees of freedom is essential
for understanding and extracting physical information from these
theories. This is usually done by a method called gauge fixing, which
is based on the assumption that a gauge-fixing condition can be found
which uniquely determines a representative gauge field on each orbit.
It was pointed out by Gribov \cite{Gr} that the standard gauge-fixing
conditions used for perturbative calculations do not in fact fix the gauge
fields uniquely: for a non-abelian gauge theory, in the Coulomb or in
the Landau gauge, there are many gauge equivalent configurations
satisfying the Coulomb or Landau
transversality condition. The existence of these {\em Gribov
copies} does not affect the results from perturbation theory, but
their elimination
could play a crucial role for non-perturbative features of
these theories.

One of the celebrated advantages of lattice gauge theories is that the
lattice provides a regularization which makes the gauge group compact,
so that the Gibbs average of any gauge-invariant quantity is
well-defined and therefore gauge fixing is, in principle,
not required. However, because of asymptotic freedom, the continuum
limit is the weak-coupling limit, and a weak-coupling expansion
requires gauge fixing. Moreover,
gauge fixing is used in smearing techniques,
and is required in order to evaluate
quark/gluon matrix elements which can be used to extract
non-perturbative results from Monte Carlo
simulations.\footnote{~For a review see \cite{Rossi}.} Thus, one is
led to consider gauge-dependent quantities on the lattice as well.
Unfortunately gauge fixing on the lattice is
afflicted by the same problem of Gribov copies encountered in the
continuum case \cite{MPR,PhF,FH,MR}.
From a numerical point of view the existence of Gribov
copies in the lattice Landau gauge is not surprising.
In fact, this gauge condition \cite{W2,MO2}
is imposed by finding a gauge transformation $\{ g (x) \}$
which brings the functional
\be
{\cal E}_{U}[ g ] \;=\;
\frac{1}{8\,V} \sum_{\mu\mbox{,}\, x} \, \tr \,
             \left[ \, \1 \, -
g(x) \; U_{\mu}(x) \; g^{\dagger}(x + e_{\mu}) \;
   \right]
\label{eq:Etomin}
\ee
to a minimum.\footnote{~This definition applies to
$SU(2)$ lattice gauge theory in $4$ dimensions.
Here $\{ U_{\mu}(x) \}$
is a given ({\em i.e.}\ fixed) thermalized lattice configuration.
See Section \ref{sec:propag} for notation and definitions.}
This functional can be
seen as the action of a spin glass model \cite{MPR},
namely a multi-dimensional system characterized by a strong
disorder, and therefore it is plausible that one finds a
large number of local minima for its energy ${\cal E}_{U}$.
 
Given the appearance of Gribov copies in numerical
studies, we need to understand their influence
on the evaluation of gauge-dependent quantities.
In fact, since usually
it is not clear how an algorithm selects among different Gribov copies
and we do not know which kind of bias they introduce on the
considered quantities ({\em Gribov noise}),
numerical results using gauge fixing
could depend on the gauge-fixing algorithm, making
their interpretation conceptually difficult.

Relatively few studies
\cite{FH,hioki,PPPTV1,HKR,BBMP,MAG,NP} have been done in order
to analyze the influence of Gribov copies
on lattice quantities.
In most cases \cite{PPPTV1,HKR} it was found that 
Gribov noise is of the same order as
the numerical accuracy of the simulations,
and that it scales down as a pure statistical error.
However, in the so-called maximally
abelian gauge,
Gribov noise seems to be quite large
and to introduce a clear bias on the number of monopoles
\cite{hioki}, and on the value of the
{\em abelian} string tension\footnote{~The string
tension, as usually defined,
is a gauge-invariant quantity. However, in
the maximally abelian gauge, the gauge field is projected
onto an abelian lattice gauge field, and the string tension
constructed in terms of this abelian field is not
gauge-invariant. For details see \cite{MAG,Debbio}.}
\cite{BBMP,MAG}.
Finally, in the case of $U(1)$
lattice gauge theories, it has been shown numerically
\cite{FH,NP} that the photon propagator is
strongly affected by Gribov noise
and that only averages taken on absolute minima of
the minimizing functional --- the $U(1)$-analogue of
${\cal E}_{U}[ g ]$ defined in eq.\
\reff{eq:Etomin} --- reproduce the theoretical predictions. 
This suggests that the same could happen for the 
gluon and the ghost propagators.

In this work we present an extensive
study of Gribov noise for gluon and ghost propagators
at four different values of $\beta$, {\em i.e.}\  
$\beta = 0\mbox{,}\, 0.8\mbox{,}\, 1.6\,$ and $\, 2.7\,$.
Although these propagators are non-gauge-invariant quantities,
the study of their infra-red behavior
provides a powerful tool for increasing our
understanding of QCD and, in particular, for gaining insight
into the physics of confinement in non-abelian gauge theories.
Notice that the first three values of $\beta$ are in the strong-coupling
regime, where the number of Gribov copies is higher \cite{MR} and
Gribov noise, if present, is probably larger and more easily
detectable.

Gluon propagators in Landau gauge
have been the subject of several numerical
studies \cite{MO2,MMS,Nakamura}.
However, in these works
Gribov copies were not taken into account.
In reference \cite{HKR} Gribov noise has been estimated
for the electric and the magnetic screening masses ---
obtained from the behavior of the gluon propagator in
the deconfined phase --- and found to be of the order
of magnitude of the numerical accuracy.
Finally, only one numerical study \cite{Ghost} is
available for the ghost propagator. Also in this case
Gribov noise was not directly analyzed.
However, since the fluctuations of the data appear to be
very small, the authors concluded that, if present, Gribov
noise should be a small effect. 
 

\section{Minimal Landau Gauge}
\label{section:MLG}

In order to get rid of the problem of spurious gauge copies,
Gribov proposed \cite{Gr} the use of additional gauge conditions. In
particular he restricted the physical configuration
space to the region
\be
\Omega \equiv \left\{\, A \, : \, \partial \cdot A \,=\,0\mbox{,}\,\,
{\cal M}\left[A\right]\,\geq\,0\, \right\} 
\label{eq:Omega}
\;\mbox{,}
\ee
where $\,{\cal M}\left[A\right]\,\equiv\,-\nabla\,\cdot\, D\left[A\right]\,$
is the Faddeev-Popov operator.
This region is clearly included in the hyperplane $\Gamma$ of
transverse configurations
({\em i.e.}\ $\partial \cdot A \,=\,0\,$)
and is delimited by the so-called {\em first Gribov horizon}, where
the smallest eigenvalue\footnote{~The Faddeev-Popov operator has a
trivial null eigenvalue, corresponding to a constant eigenvector.}
of the Faddeev-Popov operator is zero.

Nowadays we know that the region $\Omega$ is {\em not} free of Gribov copies
and that conditions stronger than those in \reff{eq:Omega}
should be imposed. In particular the physical configuration
space has to be identified with
the so-called {\em fundamental modular region} $\Lambda$, which is
defined (in the continuum)
as the set of {\em absolute} minima of the functional
\be
E_{A}[ g ]\,\equiv\,\|\,A^{(g)}\,\|^{2}\,
\equiv\,\frac{1}{2}\,\sum_{\mu\mbox{,}\,a}\,
\int\,d^{4}x\,\left\{\,\left[\,A^{(g)}\,\right]^{a}_{\mu}(x)\,\right\}^{2}
\;\mbox{.}
\label{eq:Econt}
\ee
This corresponds to selecting, on a given orbit, the configuration
closest to the origin. Of course one has to show that this functional
possesses a unique absolute minimum on each orbit.
This has been proven in \cite{STSF}.
From these works we also know
that $\Omega$ is simply the region of {\em local} minima of the functional
$E_{A}[g]$, namely it includes $\Lambda$, and
is {\em not} free of Gribov copies.
It has also been shown \cite{PvB}
that in the {\em interior} of the fundamental modular region
$\Lambda$ the absolute minima are non-degenerate.
However, on the boundary of $\Lambda$
there are degenerate absolute minima, and they have to
be identified in order to obtain a region truly free
of Gribov copies.\footnote{~This result does not contradict
Singer's no-go theorem \cite{Singer}, since the
hypothesis of continuity of the gauge-fixing condition
is dropped in this approach.}


To eliminate Gribov copies on the lattice we
can define a fundamental modular region also in this case.
To this end we can look for the absolute minimum of the
minimizing functional ${\cal E}_{U}[ g ]$ ({\em
minimal Landau gauge}), defined in \reff{eq:Etomin},
instead of looking for a generic relative minimum.
This definition is
the lattice analogue of the one used in the continuum
[see eq.\ \reff{eq:Econt}]. In fact,
it is clear that, by choosing the absolute
minimum, we select on each gauge orbit the configuration
closest to the {\em vacuum} $V_{\mu}(x) \,=\, \1\,$.
It can also be shown \cite{Z1,Z2} that the set of all
minima --- local and absolute ones --- of the functional
${\cal E}_{U}[ g ]$ is the lattice analogue of the
region $\Omega$ introduced by Gribov, namely it is the set
of transverse configurations for which the lattice Faddeev-Popov
operator
$\,{\cal M}[ U ] \equiv - \nabla\,\cdot D[ U ]\,$
is non-negative.\footnote{~Here                                        
$\nabla$ is the lattice four-divergence operator
and $D[ U ]$ is the lattice gauge-covariant derivative.}

Of course,
on a finite lattice, the existence of an absolute minimum
for ${\cal E}_{U}[ g ]$ is manifest. In fact the gauge orbit is compact
and this functional is bounded. However, from a numerical
point of view, the search for this absolute minimum
is a difficult task. The reason is
that eq.\ \protect\reff{eq:Etomin}
represents the energy of a nonlinear sigma-model
with variables $g(x)$ taking values on
a group [in our case $SU(2)$]
and, due to this non-linearity,
the search for the absolute minimum is highly non-trivial.
Moreover, there are no theoretical predictions that allow us to
distinguish a generic relative minimum from the
absolute one, except for looking at all minima of the
minimizing functional and checking that the (supposed)
absolute minimum effectively corresponds to the smallest
value of this functional.\footnote{~Recently a method to make an extrapolation
in the Gribov noise, namely to extrapolate the data in the
limit of an infinite number of gauge-fixed gauge copies,
has been introduced \cite{MAG}. An
extension of the present work implementing this extrapolation
method will be discussed in a future paper.}
The search is therefore a statistical one: enough
gauge-fixed gauge copies $C_{j}$ --- for each thermalized
configuration $C$ --- have to be produced so that the
probability to get, with a new copy, a value
of the functional ${\cal E}_{U}[ g ]$ smaller than all the previous ones is
negligible.  In particular, we need a statistical
criterion in order to decide when to stop this search. Of course, this
criterion must balance the necessity of producing enough Gribov copies and
the need to keep the computational work limited to a reasonable
amount of CPU time. We have decided to proceed in the following way.
Given a thermalized configuration $C$, we start by producing
five gauge-fixed gauge copies $C_{j}$.
The copy characterized by the smallest value for the
minimizing functional is chosen as a candidate for the
absolute minimum $C_{min}$. Then we keep doubling the number of
gauge-fixed gauge copies $C_{j}$. After each
doubling we check if, in one of the new copies,
the minimizing functional gets a value smaller than that of the supposed
absolute minimum $C_{min}$. In the case of a negative
result for this check, the search for the absolute minimum
is stopped and $C_{min}$ is used as a chosen
candidate for it. In the opposite case, the new gauge-fixed
configuration, characterized by the smallest value for the
minimizing functional, is chosen as a new candidate
for the absolute minimum $C_{min}$, and we double again
the number of Gribov copies. When $320$ copies have been produced
the search stops anyhow, and the configuration
which is the present candidate for $C_{min}$
is considered as the real absolute minimum.
So, for any configuration $C$, a minimum of $10$ and a maximum
of $320$ Gribov copies are produced.
 
Finally, we consider two different averages:
the average considering only the supposed absolute
minima (denoted by ``am''), which should
give us the result in the minimal Landau gauge;
and the average considering
only the first gauge-fixed gauge copy generated
for each configuration (denoted by ``fc''). The latter
average is the result that we would obtain if Gribov noise were not
considered.


\section{Propagators on the Lattice}
\label{sec:propag}

We consider a standard Wilson action for
$SU(2)$ lattice gauge theory in $4$ dimensions
with periodic boundary conditions.
The symbol $\,U_{\mu}(x)\in SU(2)\,$ is used
for link variables,
$\,g(x)\in SU(2)\,$ for site variables,
$e_{\mu}$ represents a unit vector in the positive $\mu$ direction,
and $ V \equiv N^{4} $
is the lattice volume. (Here we always consider lattices
with equal sizes in the four directions.)
The gauge field is defined as
\be
A_{\mu}(x) \equiv \frac{1}{2} \,
\left[ \, U_{\mu}(x) - U_{\mu}^{\dagger}(x) \, \right]
\;\mbox{.}
\ee
We also define
$A_{\mu}^{a}(x) \equiv ( 1 / 2 i )\, \tr [ A_{\mu}(x)\, \sigma^{a} ]$,
where $\sigma^{a}$ is a Pauli matrix.
Note that $ A_{\mu}^{a}(x) $ approaches
$( 1 / 2 ) [ A^{(cont.)} ]^{a}_{\mu}(x)$ in the continuum
limit, where $[ A^{(cont.)} ]^{a}_{\mu}(x) $ is the conventional
vector potential.
 
If the configuration $\{ U_{\mu}\left(x\right) \}$ is a stationary
point of the functional ${\cal E}_{U}[ g ]$,
defined in eq.\ \reff{eq:Etomin}, then \cite{W2}
the lattice divergence of $ A_{\mu}^{a}(x) $
is null, namely
\be
\left(\nabla \cdot A \right)^{a}(x)\,\equiv\,
\sum_{\mu} \, A_{\mu}^{a}(x) -
                  A_{\mu}^{a}(x - e_{\mu}) \, =
\, 0 \qquad \qquad
  \forall \; \; \; x \mbox{,} \; \; a
\;\mbox{.}
\label{eq:diverg0}
\ee
This is the lattice formulation of the usual Landau
gauge-fixing condition in the continuum.
By summing equation (\ref{eq:diverg0})
over the components $x_{\mu}$ of $x$
with $\mu \neq \nu$, for fixed $\nu$,
and using the periodicity of the lattice, it is easy
to check \cite{MO2} that if the Landau gauge-fixing condition is
satisfied then the quantities
\be
Q_{\nu}(x_{\nu}) \, \equiv \, \sum_{\mu \neq \nu} \,
     \sum_{x_{\mu}} \, A_{\nu}(x)  \qquad \qquad
  \phantom{forall} \; \; \; \nu = 1\mbox{,}\ldots\,\mbox{,} 4
\label{eq:charges}
\ee
are constant, {\em i.e.}\ independent of $x_{\nu}$.

With this notation
the lattice space-time gluon propagator is given by
\be
D_{\mu\, \nu}^{a\, b}( x - y )
\equiv \langle\,A^{a}_{\mu}(x)\, A^{b}_{\nu}(y)\,\rangle
\;\mbox{.}
\ee
To go to momentum space we can use formula (3.1a)
in reference \cite{Z1} and obtain
\ba
D(0)& \equiv& \frac{1}{12 V} \sum_{\mu\mbox{,}\,a}\,\langle\,
  \left[\,\sum_{x}\,A_{\mu}^{a}(x)\,\right]^{2} \,\rangle
\label{eq:D0def} \\
D(k) & \equiv & \frac{1}{9 V} \sum_{\mu\mbox{,}\,a}\,\langle\,
\left\{\,\left[\,\sum_{x}\,A_{\mu}^{a}(x)\,
\cos{( 2 \pi k \cdot x )}\,\right]^{2} +\,
   \left[\,\sum_{x}\,A_{\mu}^{a}(x)\,
\sin{( 2 \pi k \cdot x )}\,\right]^{2}
\, \right\} \,\rangle
\label{eq:Dkdef}
\ea
(Here $\mu$ goes from $1$ to $4$, and $k$ has components
$k_{\mu}$ given by
$k_{\mu}\,N \, \equiv \, 0\mbox{,}\,1\mbox{,}
\,2\mbox{,}\,\ldots \mbox{,}\, N - 1 $, where $N$ is the lattice
size.)
Notice that $D(0)$ is not given by $D(k)$ at $k = 0$.
The difference is due to the Landau gauge condition --- the
continuum-like condition as in equation \reff{eq:diverg0} --- which
in momentum space reads
\be
\sum_{\mu}\, p(k_{\mu}) \, {\widetilde A}_{\mu}^{a}(k)\,\equiv\,
2\,\sum_{\mu}\, \sin{\left( \, \pi \,k_{\mu} \, \right)}
\, {\widetilde A}_{\mu}^{a}(k)\, =\, 0
\;\mbox{.}
\label{eq:defp}
\ee
If $k \neq 0$ we obtain that only
three of the four Lorentz components of ${\widetilde A}^{a}(k)$
--- and therefore of $A^{a}(x)$ ---
are independent:
this explains the factor $9$ (instead of $12$) in equation (\ref{eq:Dkdef}).

The zero three-momentum-space gluon propagator
is defined as \cite{MO2,MMS}
\be
D( 0\mbox{,}\, d ) \equiv
  \frac{N}{9 V} \,\sum_{t\mbox{,}\, i\mbox{,}\,a}\,
\langle \, {\cal A}_{i}^{a}( t )\,
{\cal A}_{i}^{a}( t + d ) \,\rangle
\;\mbox{,}
\label{eq:D0d}
\ee
where invariance under space-time translations is used in
order to improve statistics.\footnote{~The normalization in
\protect\reff{eq:D0d} has been chosen so that
the first term in equations \protect\reff{eq:D0} and
\protect\reff{eq:PhiD0}
can be directly related to the zero-momentum gluon
propagator in eq.\ \protect\reff{eq:newD0}.} Here $ i $ goes
from $1$ to $3$ (the three directions orthogonal to the
time direction, {\em i.e.} $ i = 4$) and
\be
{\cal A}_{i}^{a}(t) \equiv \sum_{\bvx}\,A_{i}^{a}( \bvx\mbox{,}\, t )
\;\mbox{.}
\ee
Notice that, if we denote by $\phi^{a}_{i}$ the zero-momentum component
of the gluon field $A_{i}^{a}(x)$, namely
\be
\phi^{a}_{i} \equiv \frac{1}{V} \,\sum_{x}\,
      A^{a}_{i}(x)
\label{eq:defphi}
\ee
(where i = 1\mbox{,} 2\mbox{,} 3\mbox{,} 4),
then we can write
\be
{\cal A}_{i}^{a}(t) \equiv \frac{V}{N}\,
   \phi^{a}_{i}\,+\, [ {\cal A}^{(0)} ]_{i}^{a}(t)
\ee
and the relation
\be
\sum_{t}\, [ {\cal A}^{(0)} ]_{i}^{a}(t) \,=\, 0
\ee
holds for any $i$ and $a$.
Thus we obtain, for the zero three-momentum-space gluon propagator,
the expression
\ba
D( 0\mbox{,}\, d ) & = & \frac{V}{9} \, \sum_{i\mbox{,}\,a}\,
\langle \, ( \phi^{a}_{i} )^{2} \,\rangle\,+\,
  \frac{N}{9 V} \,\sum_{t\mbox{,}\, i\mbox{,}\,a}\,
\langle \, [ {\cal A}^{(0)} ]_{i}^{a}( t )\,
[ {\cal A}^{(0)} ]_{i}^{a}( t + d ) \,\rangle
\label{eq:D0} \\
& \equiv & \Phi^{2} 
\,+\,
   D^{(0)}( 0\mbox{,}\, d )
\label{eq:PhiD0}
\;\mbox{.}
\ea
Clearly, the contribution of the constants $\phi^{a}_{i}$
to the propagator is considerably enhanced by the volume
factor in the first term.
The effect of the positive constant $\Phi^{2}$ in
eq.\ \reff{eq:PhiD0}
on masses obtained in numerical lattice simulations
has been recently pointed out in reference \cite{Mitr}.
Here we evaluate both the total zero three-momentum
gluon propagator, as defined in \reff{eq:D0d}, and
the propagator $D^{(0)}( 0\mbox{,}\, d )$ defined in
\reff{eq:D0} and \reff{eq:PhiD0}.

Let us observe that a nonzero value for the constants
$\phi_{i}^{a}$ is related to the use of periodic boundary
conditions and to the finiteness of the volume. In fact,
with free boundary conditions, these constants
are identically null, even on a finite lattice,
after the Landau gauge condition
is imposed \cite{SZ}. Moreover, in the periodic case,
it has been proven \cite{Z1,Z4}
that these constants must go to zero as the volume increases;
more exactly, one obtains $\phi_{i}^{a} \ltapprox 1/N $ as the
lattice size $N$ goes to infinity. This bound
is a consequence only of the positivity
of the Faddeev-Popov matrix in the lattice Landau gauge.
Thus, a finite nonzero value of $\phi_{i}^{a}$
is a lattice artifact due to finite volume and
translational invariance. Note that this does not imply that
$\Phi^{2}$ should be
zero in the infinite volume. However,
Zwanziger proved \cite{Z1,Z2,Z4} that, in the infinite-volume
limit, the zero four-momentum gluon propagator $D(0)$, which
can be written as\footnote{~Notice that here $\mu$ goes
from $1$ to $4$, while in eq.\ \protect\reff{eq:D0} the index
$i$ goes from $1$ to $3$.}
\be
D(0) \, = \, \frac{V}{12} \sum_{\mu\mbox{,}\,a}\,\langle\,
    ( \phi^{a}_{\mu} )^{2} \,\rangle
\label{eq:newD0}
\;\mbox{,}
\ee
and is clearly related to $\Phi^2$,
is less singular at momentum $k = 0$ than $k^{-2}$
and that, very likely, it does vanish as rapidly as $k^{2}$.
Thus, the infinite-volume limit of $\Phi^{2}$
is related to the infra-red behavior of the gluon propagator,
a problem still not settled.

Finally, following \cite{Ghost,Z2}, we define the space-time ghost
propagator as
\be
G(x - y)\,\delta^{a b}\,\equiv\,\langle\,
     \left(\,{\cal M}^{- 1}\,\right)^{a\, b}_{x\, y}[ U ]
       \,\rangle
\;\mbox{.}
\ee
Going to momentum space and taking the trace in color space,
we obtain
\be
G(k)\, = \,\frac{1}{3 V}
\sum_{x\mbox{,}\, y} e^{- 2 \pi i \, k \cdot (x - y)}\,
\sum_{a}\,\langle\,
     \left(\,{\cal M}^{- 1}\,\right)^{a\,a}_{x\, y}[ U ]
       \,\rangle
\label{eq:Gk}
\;\mbox{.}
\ee
When the lattice Landau gauge is fixed,
the Faddeev-Popov matrix
is symmetric and semi-positive-definite \cite{Z2},
and we can write
\be 
{\cal M}[ U ]\,=\,
- \left( \nabla\,\cdot D[ U ] \right) \,=\,
- \left( D[ U ] \,\cdot \nabla \right)
\;\mbox{.}
\ee
(For an explicit definition of the matrix
${\cal M}^{a\, b}_{x\, y}[ U ]$
see eq.\ (B.18) in reference \cite{Z2}.)
In particular, this matrix is positive-definite in the subspace
orthogonal to constant vectors. Therefore, it can be inverted by using
a standard conjugate-gradient method (CG), provided that we work
in this subspace. Moreover, since we use this method
in order to evaluate the Faddeev-Popov propagator, we need
to impose this restriction in a way that preserves translational
invariance for the lattice. This is not a problem if the source
$\psi^{a}(x)$ and the initial guess of the solution
for the CG-method have zero constant mode \cite{Ghost,Athesis}.
As a source we have used
\be
\psi^{a}(x) \,=\,\delta^{a c} \, e^{- 2 \pi i \, k \cdot x}
\;\mbox{,}
\ee
where $c$ has a fixed value ($1\mbox{,}\,2$ or $3$) and
$k \neq 0$. In this way,
the condition $\sum_{x}\, \psi^{a}(x)\,=\,0$ is
automatically imposed, and at the same time we save computer time by
evaluating ``half'' of the Fourier transform in eq.\ \reff{eq:Gk}.

\section{Numerical Simulations and Results}

To thermalize
the gauge configuration $\{U_{\mu}(x)\}$
at a fixed value of the coupling
$\,\beta\,$
we use a hybrid overrelaxed (HOR) algorithm \cite{BW}.
This algorithm is able to speed up
numerical simulations in the
case of pure lattice gauge theory: in fact,
an overall efficiency gain of
up to a factor $10$ has been observed \cite{DFor}
[for the $SU(2)$ case] with respect to the heat-bath algorithm,
and its {\em dynamic critical exponent}\footnote{~For a
definition of dynamic critical exponent see for example \cite{S}.} has been
found \cite{zhybrid} to be equal to $1.0(1)$
[again in the $SU(2)$ case].
The HOR algorithm is also
easy to be implemented:
$m$ microcanonical (or energy-conserving) update sweeps
are done, followed by one standard local ergodic update
(heat-bath sweep) of the lattice. In order to
optimize the efficiency of the heat-bath code, we implement two
different $SU(2)$ generators
(methods 1 and 2 described in Appendix A of \cite{EFGS}, with
$h_{cutoff} = 2$).
In our case we did {\em not} tune
the value of $m$; following \cite{zhybrid2} we
fixed $m = N / 2$.
However, for all the pairs
$(\beta\mbox{,}\, N)$, we evaluated
the {\em integrated autocorrelation time}\footnote{~For a definition
see \cite{S}. To evaluate the integrated autocorrelation time we use
an automatic windowing procedure \cite{S} with two different window
factors ($6$ and $15$). We also employ a method \cite{zhybrid2}
based on a comparison between the naive statistical error with a 
jack-knife binning error \cite{W2}.
In all cases we checked that these three estimates are in agreement.}
for the Wilson loops $\,W(l,l)\,$ with
$\,l\,=\, 1\mbox{,}\,2\mbox{,}\,4\mbox{,}\,\ldots\mbox{,}\,
N / 2\,$, and for the Polyakov loop $P\,$ (denoted respectively by
$\,\tau_{int,W_{l}}\,$ and $\,\tau_{int,P}\,$). For all pairs and
observables, except for $P$ at $\beta = 2.7$, we
obtained $\tau_{int} \ltapprox 1$.
Note that $\tau_{int} \,=\, 0.5$ indicates
that two successive configurations generated in the Monte
Carlo simulation are independent.
The situation is different
at $\beta = 2.7$, where there is clearly a slow
mode associated with the Polyakov loop
$P$, a result in agreement with the analysis
in reference \cite{GP}. At any rate, also in this case we
obtained $\,\tau_{int,P} \ltapprox 20$ for all lattice sizes.
Since, for all pairs $(\beta\mbox{,}\, N)$
and for all quantities, the
integrated autocorrelation time is much smaller than
the number of sweeps between two consecutive configurations
used for evaluating the propagators (see Table \ref{Table:thermalization}),
we can conclude that these configurations are essentially
statistically independent. For the lattice $8^{4}$
at $\beta = 2.7$ we also made a comparison between the heat-bath and the
HOR algorithm. We obtained $\tau_{int,P}\,=\,225 \pm 68$
in the first case and $\tau_{int,P}\,=\,12.5 \pm 1.2$ in the
second one (see \cite{Athesis} for details). Since,
in our case,
the deterministic update of a link is almost five times faster
than the heat-bath one, and ``one sweep'' of the lattice
for the $HOR$ algorithm means
$m = 4$ microcanonical sweeps followed by one heat-bath sweep,
we gain a real ({\em i.e.}\ CPU-time) factor of order $10$,
in agreement with \cite{DFor}.

For the numerical gauge fixing we use
the so-called {\em stochastic overrelaxation} algorithm \cite{FG},
which has dynamic critical exponent $z$ equal to $1$ (see \cite{CM}).
In all our simulations we stop the gauge fixing when the condition
\be
\frac{1}{V} \sum_{x\mbox{,}\, a} \, \Big[
    \left( \nabla \cdot A \right)^{a}(x) \Big]^{2} \, \leq \,
     10^{- 12}
\label{eq:divergenza2}
\ee
is satisfied. This
is equivalent \cite{CM} to fixing the minimizing functional up to
about one part in $10^{12}$,
and it is sufficient in order to distinguish two different
lattice Gribov copies \cite{MPR,PhF,FH,MR}.
In the final gauge-fixed configuration
we also evaluate
\be
Q \, \equiv \, \frac{1}{12\,N} \, \sum_{\nu}\,
   \sum_{x_{\nu}\mbox{,}\, a}  \,
    \left[ \,  Q_{\nu}^{a}(x_{\nu}) - {\widehat Q}_{\nu}^{a}  \,
      \right]^{2} \, \left[ {\widehat Q}_{\nu}^{a} \right]^{- 2}
\label{eq:Qdefinizione}
\;\mbox{,}
\ee
where $\,
{\widehat Q}_{\nu} \,
\equiv \, (\, 1 / N\, ) \, \sum_{x_{\nu}} \,
            Q_{\nu}(x_{\nu})\,$,
and $Q_{\nu}(x_{\nu})$ has been defined in
eq.\ \reff{eq:charges}.
This quantity
should be zero when the configuration is gauge-fixed, and it
is a good estimator of the quality of the gauge fixing.
As in reference \cite{CM}, we found
that the stochastic overrelaxation algorithm is
very efficient in fighting {\em critical slowing-down} \cite{S}, and
in making the quantities $Q_{\nu}(x_{\nu})$ constant.

In Table \ref{Table:thermalization} we report, for each
pair $(\beta\mbox{,}\, N)$, the parameters
used for the simulations.
Overall, we have
considered about $2000$ configurations (gauge orbits)
and $40,000$ Gribov copies.
In all our runs we have started from a randomly chosen
lattice gauge configuration.

From tests conducted at
lattice sizes up to $8^{4}$, we noticed a strong
violation of rotational invariance, as is expected
for small lattices and non-improved actions.
Therefore, we decided to evaluate the four-momentum-space
gluon and ghost propagators by considering only values of
$k$ with three of the four components equal to zero,
namely $k = (0\mbox{,}\, 0\mbox{,}\, 0\mbox{,}\, k_{4})$.

Finally, for the ghost propagator at $\beta = 1.6$,
data have been collected only up to the volume $16^{4}$, due to
limitations on the memory of the available computers.

Computations were performed on
several IBM RS-6000/250--340 workstations
at New York University.

\subsection{Gluon Propagator}

In Table \ref{Table:gluon}
we show the results (averages ``am'' and ``fc'')
for the four-momentum-space
gluon propagator $D(k)$ as a function of 
$N k_{4}$.
The data corresponding to the minimal Landau gauge (absolute
minima) are in complete agreement, within
statistical errors, with those obtained in a generic
Landau gauge. This happens even at $\beta = 0$, where the number of Gribov
copies is very large (see Table \ref{Table:thermalization}).
Clearly for the zero three-momentum-space gluon propagator
the result is the same:
Gribov noise, if present, seems to be irrelevant
compared to the statistical fluctuations. This is
the case for both propagators, namely $D( 0\mbox{,}\, d )$ 
defined in eq.\ \reff{eq:D0d} and
$D^{(0)}( 0\mbox{,}\, d )$ defined in eq.\
\reff{eq:D0} and \reff{eq:PhiD0}.

However, a few interesting observations can be made
about these data.

First of all, from Table \ref{Table:gluon}
we can see that, in the strong-coupling regime, the
infra-red behavior of the gluon propagator is completely different
from that at $\beta = 2.7$, where it grows very fast
as the momentum goes to zero. In particular, at $\beta = 0$,
this propagator is clearly {\em decreasing} as $k_{4}$ goes to zero.
At $\beta = 0.8$ we see an almost constant propagator.
Finally, at $\beta = 1.6$ the propagator is increasing
as $k_{4}$ goes to zero for the lattice size $N = 8$, while
it is constant or slightly decreasing with $k_{4}$
for $N = 16$ and $N = 24$. Also the behavior of
zero-momentum gluon propagator $D(0)$ as
a function of the lattice size $N$ is very $\beta$-dependent.
In fact, at $\beta = 0$, it decreases as $N$ increases, and from
our data it is not clear if its value would be zero,
or a finite nonzero constant, in the limit
of $N$ going to infinity. On the contrary, at $\beta = 2.7$,
this quantity is increasing
with the lattice size. Finally, at the two intermediate
values of $\beta$ we see a very small volume dependence.

\begin{figure}[p]
\begin{center}
\vspace*{0cm} \hspace*{-0cm}
\epsfxsize=0.9\textwidth
\epsffile{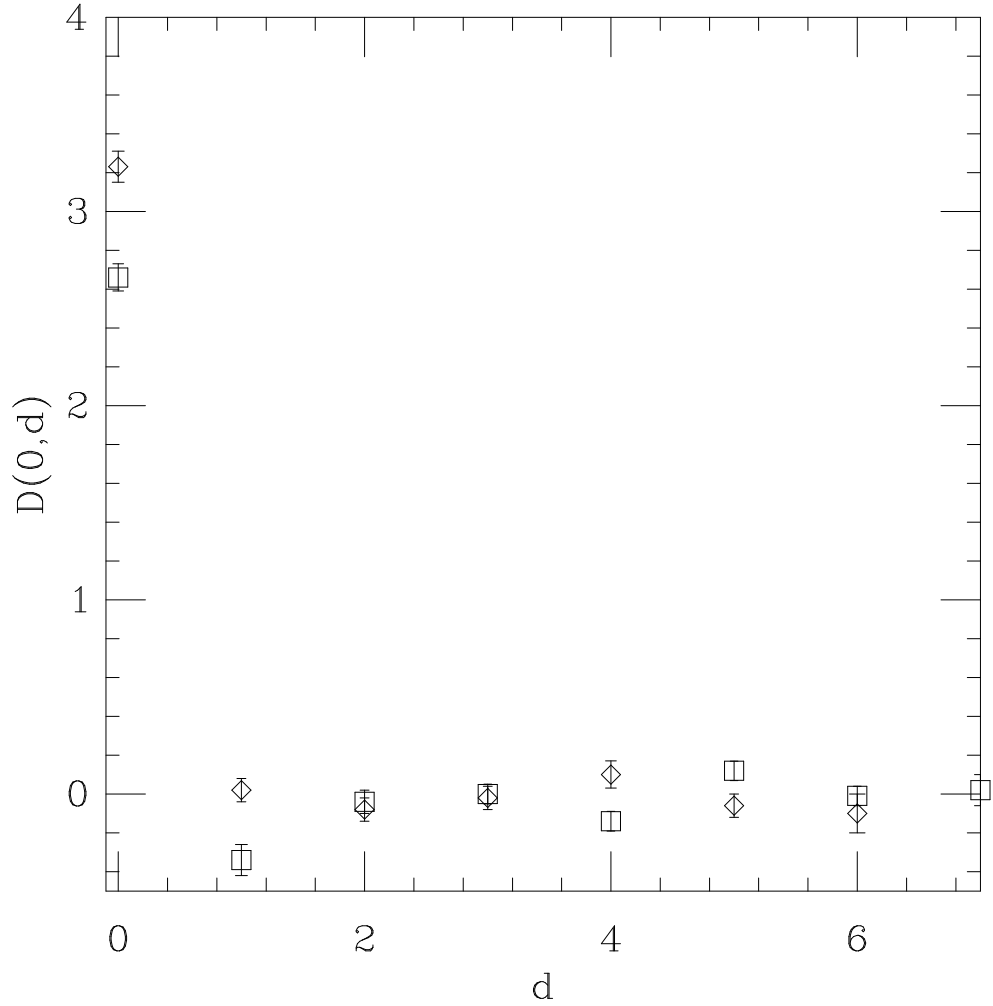} \\
\end{center}
\vspace{-0.5cm}
\caption{~The zero three-momentum-space
          gluon propagator $D( 0\mbox{,}\, d )$
          [see eq.\ \protect\reff{eq:D0d}] as a function of the
          time separation $d$. Since we use periodic boundary conditions,
          only data for $d \leq N/2$
          are reported here.
          Data correspond to
          $V = 14^{4}$ at $\beta = 0$ ($\Box$),
          and $V = 12^{4}$ at $\beta = 0.8$ ($\Diamond$).
          Here we consider the ``fc'' statistics
          (for a definition see the last paragraph in
          Section \protect\ref{section:MLG}.) Error bars
          are one standard deviation.
}
\label{fig:gluon3_fin}
\end{figure}

Related to these observations is the fact that,
in the strong-coupling regime,
the gluon propagator $D( 0\mbox{,}\, d )$ is
not always positive (see Figure \ref{fig:gluon3_fin}).
This is a manifest violation
of reflection positivity \cite{Z1,newpaper}.

\begin{figure}[p]
\begin{center}
\vspace*{0cm} \hspace*{-0cm}
\epsfxsize=0.9\textwidth
\epsffile{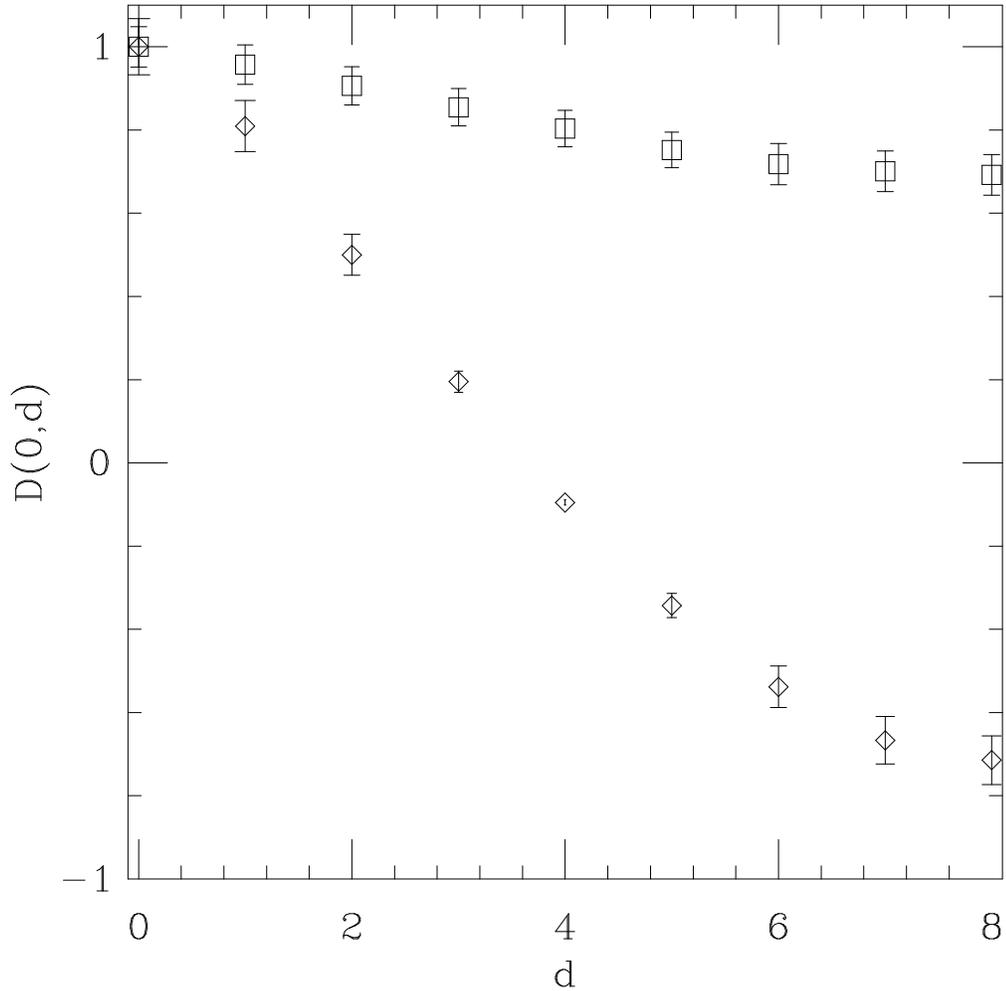}  \\
\end{center}
\vspace{-0.5cm}
\caption{~Comparison between the zero three-momentum-space
          gluon propagators $D( 0\mbox{,}\, d )$
          ($\Box$)
          and $D^{(0)}( 0\mbox{,}\, d )$
          ($\Diamond$)
          [see eqs.\ \protect\reff{eq:D0d}
          and \protect\reff{eq:D0}] as a function of the
          time separation $d$.
Since we use periodic boundary conditions,
only data for $d \leq N/2$ are reported here.
The data, normalized to one at the zero time-separation,
correspond to $V = 16^{4}$ and $\beta = 2.7$.
Here we consider the ``fc'' statistics
          (for a definition see the last paragraph in
          Section \protect\ref{section:MLG}.)
Error bars are one standard deviation.
}
\label{fig:gluon3_0_fin}
\end{figure}

Finally, looking at Figure \ref{fig:gluon3_0_fin}, it is clear
that the positiveness of $D( 0\mbox{,}\, d )$
at $\beta = 2.7$ is only due to the large value
of the positive constant in eq.\ \reff{eq:PhiD0}, as
predicted in reference \cite[Fig.\ 1]{Mitr}.

\subsection{Ghost Propagator}

Results for the ghost propagator are reported
in Table \ref{Table:ghost}. In this case
it can be noticed that the data obtained
in the minimal Landau gauge (average ``am'') are
constantly smaller than or equal to the corresponding
``fc''-data.\footnote{~This is not the case
only for $\beta = 1.6 \mbox{,} \, N = 16$ and
$N k_{4} = 1$.} Moreover, when the statistics are good,
this difference is larger than
the numerical uncertainties, especially at small
values of $N k_{4}$. The only case in which this
does not happen is $\beta = 2.7$. However, in this case, even
for $N = 16$, almost no Gribov copies were
produced, and therefore
we cannot expect a difference between the two sets of
data (see Table \ref{Table:thermalization}).
At any rate, Gribov noise
can clearly be observed for the three strong-coupling
values of $\beta$.

This result can be qualitatively explained.
In fact, as we said in Section \ref{section:MLG},
the smallest non-trivial eigenvalue $\lambda_{min}$
of the Faddeev-Popov operator ${\cal M}[ U ]$ goes to zero as the
first Gribov horizon ({\em i.e.}\ the boundary of the region $\Omega$)
is approached. So, $\lambda_{min}$ can
be interpreted as a sort of distance between the configuration
and $\partial \Omega$. Since the fundamental modular region
$\Lambda$ is included in $\Omega$, the absolute minimum,
which belongs to $\Lambda$,
should be ``farther'' from the the boundary of $\Omega$ than a
generic relative minimum.
Thus, the absolute minimum configuration
should correspond to a value of $\lambda_{min}$ larger --- in average ---
than the value obtained in a generic relative minimum.\footnote{~A
numerical study of this eigenvalue and its
dependence on Gribov copies has been done in reference \cite{Athesis}
and will be reported in a future paper \cite{future2}. Results seem
to confirm the scenario described here.} This would
imply [see eq.\ \reff{eq:Gk}] a ghost propagator
smaller (in average) at the absolute minimum, as observed in 
Table \ref{Table:ghost}.

\begin{figure}[p]
\begin{center}
\vspace*{0cm} \hspace*{-0cm}
\epsfxsize=0.9\textwidth
\epsffile{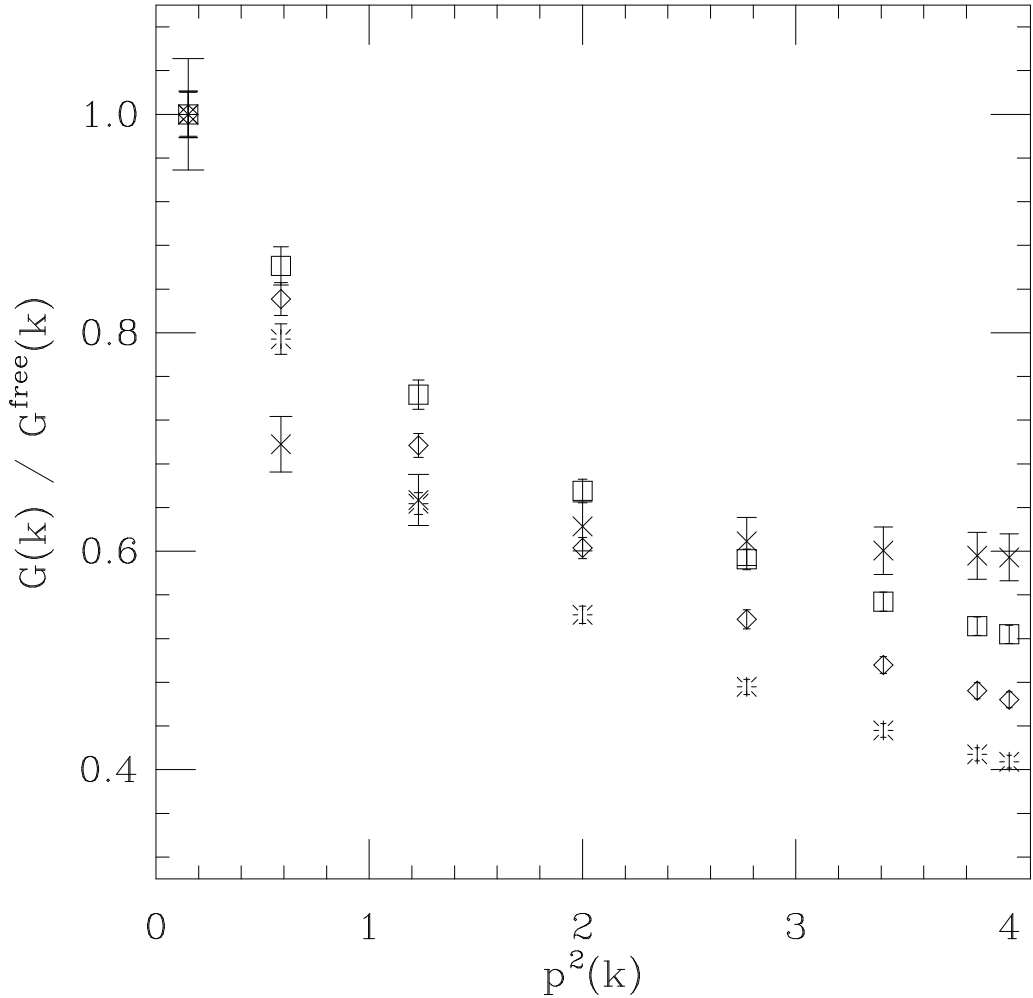}  \\
\end{center}
\vspace{-0.5cm}
\caption{~Plot of the
four-momentum-space ghost
          propagator $G(k)$ [see eq.\ \protect\reff{eq:Gk}]
          divided by the corresponding free propagator
          $G^{free}(k)$ [see eq.\ \protect\reff{eq:Gfree}]
          as a function of $p^{2}(k)$ [see eq.\
          \protect\reff{eq:defp}].
In all our runs we set $k = (0\mbox{,}\,
          0\mbox{,}\, 0\mbox{,}\, k_{4})$.
Since we use periodic boundary conditions,
only data for $k_{4} \leq 1/2$ are reported here.
The data, normalized to one at the smallest
nonzero momentum $k_{4} = 1$,
correspond to $V = 16^{4}$ at
$\beta = 0$ ($\Box$),
$\beta = 0.8$ ($\Diamond$),
$\beta = 1.6$ ($\ast$),
$\beta = 2.7$ ($\times$).
Here we consider the ``am'' statistics
          (for a definition see the last paragraph in
          Section \protect\ref{section:MLG}.)
Error bars are one standard deviation.
}
\label{fig:ghost2}
\end{figure}

Recently, Zwanziger proposed \cite{Z2}
a modification of the $SU(N)$-Yang-Mills action
which effectively constrains the functional integral to the fundamental
modular region $\Lambda$ in minimal Landau gauge.
The lattice partition function of this model can
be directly evaluated, by the saddle-point method,
in the thermodynamic limit.
From this theory, it is possible to conclude \cite{Z2}
that, in four Euclidean
dimensions, the lattice ghost propagator has a dipole singularity
at zero momentum.
In order to check the infra-red behavior of the ghost
propagator we divided $G(k)$
(data in Table \ref{Table:ghost}, average ``am'')
by the corresponding free propagator
\be
G^{free}(k) \equiv [\, 4\, \sum_{\mu}\,
          sin^{2}{( \pi k_{\mu} )} \, ]^{- 1} \equiv  1 / p^{2}(k)
\label{eq:Gfree}
\;\mbox{.}
\ee
In Figure \ref{fig:ghost2} we plot the results
for the four different values of $\beta$ and
lattice size $N = 16$. (Similar results can be obtained for
different lattice sizes). 
From these data it is clear that $G(k)$ diverges faster than
$1 / p^{2}(k)$ as $k$ goes to zero. In order to analyze this
singularity we have tried a fit of the data in Table \ref{Table:ghost}
(average ``am'') using the Ansatz
\be
G(k) \,=\, a\, p^{-2}(k)\, +\, b\, p^{-4}(k) \,+\, \ldots
\;\mbox{,}
\ee
as in reference \cite{Ghost}. In all cases we have found that
the fit with $b$ set to zero is very poor. When the dipole
singularity is included, the fit improves and we
have $a \gtapprox b$. However, for large lattice sizes and
small momenta, the fit always overestimates the
numerical values of the data. This is in agreement with results
in reference \cite{Ghost}, and suggests a singularity at zero momentum
smaller than $p^{-4}(k)$. As a second attempt we tried a
fit\footnote{~In both cases the fit has been done using
a {\tt mathematica} program implementing the
singular-value-decomposition method (see for example
\cite{recipes}).} to the Ansatz
\be
G(k) \,=\, a\, [\,p^{-2}(k)\,]^{s}
\label{eq:defs}
\;\mbox{.}
\ee
In this case the fit is in general quite good.
Finally, in order to probe the small-momentum limit,
we evaluate the exponent $s$ in eq.\ \reff{eq:defs}
using only data corresponding to the two smallest
momenta available for each lattice size. From the results, 
reported in Table \ref{Table:s2},
it can be noticed that $s$ decreases as the lattice size
increases, while it increases
as $\beta$ increases. The last result is also clear
in Figure \ref{fig:ghost2}.

\section{Conclusions}

Our data show absence of Gribov noise for the
gluon propagator and a nonzero Gribov noise for the ghost propagator.
In the latter case, the effect is small but clearly detectable
for the values of $\beta$ in the strong-coupling region.
The fact that this noise is not observable at $\beta = 2.7$
seems to us to be related only to the small volumes considered here.
Of course this hypothesis should be checked numerically.
This is, at the moment, beyond the limits of our computational resources.
We stress that this is the {\em first} case of
evidence of Gribov noise in a ``truly'' non-abelian
simulation. To date, in fact, Gribov noise was observed only
in simulations for the $U(1)$ case \cite{FH,NP}, or using maximally
abelian gauge \cite{hioki,BBMP,MAG}.

As for the infra-red behavior of these two propagators, the data
for the ghost propagator show a pole ``between'' the
zeroth-order perturbative behavior $p^{-2}(k)$ --- valid at
large momenta ---
and the $p^{-4}(k)$ singularity predicted in \cite{Z2}.
However, in this case, the volumes used do not really allow
an analysis of the small-momentum limit, and this analysis is
complicated by the fact that the ghost
propagator cannot be evaluated at zero momentum.
For the gluon propagator, at least
in the strong-coupling regime, the data are quite interesting,
showing a propagator decreasing as the momentum goes to zero.
This anomalous behavior, predicted in \cite{Gr,Z1,Z2}, is still
observable at $\beta = 1.6$, if large volumes are considered.
Related to this behavior is the violation
of reflection positivity observable in
Figure \ref{fig:gluon3_fin}.
Finally, our data show that the behavior of the
zero three-momentum-space gluon propagator is strongly
affected by the zero-momentum modes of the gluon field \cite{Mitr}.
We think that these results
deserve a more accurate numerical analysis which will be presented
in a separate paper \cite{future}.

\section*{Acknowledgements}

I am indebted to D.Zwanziger for suggesting
this work to me. I also would like to thank him,
G.Dell'Antonio, T.Mendes, V.K.Mitryushkin, S.Petrarca,
M.Schaden and B.Taglienti for valuable discussions
and suggestions, and Ph.\ de Forcrand for e-mail
correspondence.

Part of this work has been done
at the University of Rome ``Tor Vergata'';
I thank the Physics Department,
and in particular R.Petronzio and the APE
group, for the hospitality.

I also thank for the hospitality
the Physics Department
of the University of Bielefeld and the Center
for Interdisciplinary Research (ZiF) at Bielefeld, where
this work has been finished.



\clearpage

\begin{table}
%
\hspace*{-1.0cm}
\protect\footnotesize
\begin{center}
\begin{tabular}{|| c | c | c | c | c | c | c | c ||}
\hline
\hline
$ \beta $ & $ N $ & config.\ & th.\ sweeps & sweeps & $p$ & copies &
$ \mbox{fc} \neq \mbox{am} $ \\ 
\hline
$ 0.0 $ & $  4 $ & $ 400 (304) $ & $ 10 $ & $ 10 $ & $ 0.84 $ & $ 4220 (1127) $ & $ 159 $ \\ 
\hline
$ 0.0 $ & $  6 $ & $ 200 (200) $ & $ 20 $ & $ 10 $ & $ 0.875 $ & $ 4060 (3121) $& $ 167 $ \\ 
\hline
$ 0.0 $ & $  8 $ & $ 200 (200) $ & $ 10 $ & $ 10 $ & $ 0.915 $ & $ 6120 (6071) $& $ 179 $ \\ 
\hline
$ 0.0 $ & $ 10 $ & $ 100 (100) $ & $ 20 $ & $ 10 $ & $ 0.945 $ & $ 3850 (3840) $& $ 89 $ \\ 
\hline
$ 0.0 $ & $ 12 $ & $  80 (80) $ & $ 20 $ & $ 10 $ & $ 0.945 $ & $ 1940 (1936) $ & $ 71 $\\ 
\hline
$ 0.0 $ & $ 14 $ & $  20 (20) $ & $ 20 $ & $ 10 $ & $ 0.95  $ & $  590 (590) $ 
& $ 19 $ \\ 
\hline
$ 0.0 $ & $ 16 $ & $  10 (10) $ & $ 20 $ & $ 10 $ & $ 0.96  $ & $  510 (506) $
& $ 9 $ \\ 
\hline
\hline
$ 0.8 $ & $  8 $ & $ 200 (200) $ & $ 1000 $ & $ 100 $ & $ 0.875 $ & $ 6950 (6627) $ & $ 177 $ \\ 
\hline
$ 0.8 $ & $ 12 $ & $  22 (22) $ & $ 1375 $ & $ 125 $ & $ 0.9   $ & $ 1240 (1239) $ & $ 20 $ \\ 
\hline
$ 0.8 $ & $ 16 $ & $   9 (9) $ & $ 1100 $ & $ 100 $ & $ 0.94  $ & $  140 (140) $& $ 8 $ \\ 
\hline 
\hline
$ 1.6 $ & $  8 $ & $ 100 (100) $ & $ 1100 $ & $ 100 $ & $ 0.865 $ & $ 1910 (1545) $ & $ 85 $ \\ 
\hline
$ 1.6 $ & $ 16 $ & $  20 (20) $ & $ 1650 $ & $ 150 $ & $ 0.92  $ & $  910 (905) $  & $ 19 $\\ 
\hline
$ 1.6 $ & $ 24 $ & $   7 (7) $ & $ 1650 $ & $ 150 $ & $ 0.91  $ & $  290 (290) $ & $ 6 $ \\ 
\hline
\hline
$ 2.7 $ & $  8 $ & $ 400 (47) $ & $ 2200 $ & $ 200 $ & $ 0.8   $ & $ 4020 (494) $ & $ 17 $\\ 
\hline
$ 2.7 $ & $ 12 $ & $ 158 (25) $ & $ 2200 $ & $ 200 $ & $ 0.85  $ & $ 1600 (208) $ & $ 11 $\\ 
\hline
$ 2.7 $ & $ 16 $ & $ 100 (26) $ & $ 2200 $ & $ 200 $ & $ 0.875 $ & $ 1000 (157) $ & $ 13 $ \\ 
\hline 
\hline
\end{tabular}
\end{center}
\caption{~The pairs $(\beta\mbox{,} N)$ used
         for the simulations, the number of
         configurations (in brackets, the number of
         configurations for which Gribov copies have been
         found), the number of sweeps used for thermalization,
         the number of sweeps between two consecutive
         configurations used for collecting our data,
         the parameter $p$ used by the
         stochastic overrelaxation algorithm, the total
         number of Gribov copies produced (in brackets, the total
         number of different Gribov copies), and the number of
         times the first copy produced was not the absolute
         minimum copy.
}
\label{Table:thermalization}
\vspace*{0.3cm}
\end{table}

\clearpage

\begin{table}
\addtolength{\tabcolsep}{-1.0mm}
\hspace*{-1.0cm}
\protect\tiny 
\begin{center}
\begin{tabular}{|| c | c | c | c | c | c |
 c | c | c | c | c | c ||}
\hline
\hline
$ \beta $ & $ N $ & stat. & $ 0 $ &
$  1 $ & $  2 $ & $  3 $ & $  4 $ & $  5 $ & $  6 $ & $  7 $ & $  8 $ \\
\hline
\hline
$ 0.0 $ & $ 4 $ & am & $ 0.208 (0.004)
 $ & $ 0.218 (0.003)
 $ & $ 0.226 (0.006)
 $ & $  $ & $  $ & $  $ & $  $ & $  $ & $  $ \\ \hline 
$ 0.0 $ & $ 4 $ & fc & $ 0.207 (0.004)
 $ & $ 0.221 (0.003)
 $ & $ 0.229 (0.005)
 $ & $  $ & $  $ & $  $ & $  $ & $  $ & $  $ \\ \hline 
$ 0.0 $ & $ 6 $ & am & $ 0.185 (0.006)
 $ & $ 0.189 (0.004)
 $ & $ 0.214 (0.005)
 $ & $ 0.219 (0.007)
 $ & $  $ & $  $ & $  $ & $  $ & $  $ \\ \hline 
$ 0.0 $ & $ 6 $ & fc & $ 0.183 (0.006)
 $ & $ 0.198 (0.004)
 $ & $ 0.215 (0.005)
 $ & $ 0.222 (0.007)
 $ & $  $ & $  $ & $  $ & $  $ & $  $ \\ \hline 
$ 0.0 $ & $ 8 $ & am & $ 0.162 (0.005)
 $ & $ 0.174 (0.004)
 $ & $ 0.196 (0.004)
 $ & $ 0.209 (0.005)
 $ & $ 0.209 (0.006)
 $ & $  $ & $  $ & $  $ & $  $ \\ \hline 
$ 0.0 $ & $ 8 $ & fc & $ 0.160 (0.005)
 $ & $ 0.177 (0.004)
 $ & $ 0.186 (0.004)
 $ & $ 0.209 (0.005)
 $ & $ 0.211 (0.006)
 $ & $  $ & $  $ & $  $ & $  $ \\ \hline 
$ 0.0 $ & $ 10 $ & am & $ 0.154 (0.007)
 $ & $ 0.168 (0.005)
 $ & $ 0.195 (0.006)
 $ & $ 0.211 (0.007)
 $ & $ 0.202 (0.007)
 $ & $ 0.22 (0.01)
 $ & $  $ & $  $ & $  $ \\ \hline 
$ 0.0 $ & $ 10 $ & fc & $ 0.157 (0.007)
 $ & $ 0.172 (0.006)
 $ & $ 0.196 (0.006)
 $ & $ 0.203 (0.007)
 $ & $ 0.214 (0.007)
 $ & $ 0.22 (0.01)
 $ & $  $ & $  $ & $  $ \\ \hline 
$ 0.0 $ & $ 12 $ & am & $ 0.160 (0.007)
 $ & $ 0.147 (0.005)
 $ & $ 0.186 (0.007)
 $ & $ 0.194 (0.007)
 $ & $ 0.203 (0.008)
 $ & $ 0.205 (0.007)
 $ & $ 0.21 (0.01)
 $ & $  $ & $  $ \\ \hline 
$ 0.0 $ & $ 12 $ & fc & $ 0.148 (0.007)
 $ & $ 0.169 (0.006)
 $ & $ 0.181 (0.007)
 $ & $ 0.192 (0.006)
 $ & $ 0.214 (0.008)
 $ & $ 0.210 (0.007)
 $ & $ 0.22 (0.01)
 $ & $  $ & $  $ \\ \hline 
$ 0.0 $ & $ 14 $ & am & $ 0.15 (0.01)
 $ & $ 0.17 (0.01)
 $ & $ 0.18 (0.01)
 $ & $ 0.19 (0.01)
 $ & $ 0.23 (0.02)
 $ & $ 0.23 (0.02)
 $ & $ 0.23 (0.02)
 $ & $ 0.24 (0.03)
 $ & $  $ \\ \hline 
$ 0.0 $ & $ 14 $ & fc & $ 0.15 (0.01)
 $ & $ 0.14 (0.01)
 $ & $ 0.18 (0.02)
 $ & $ 0.19 (0.01)
 $ & $ 0.18 (0.01)
 $ & $ 0.24 (0.02)
 $ & $ 0.25 (0.02)
 $ & $ 0.19 (0.02)
 $ & $  $ \\ \hline 
$ 0.0 $ & $ 16 $ & am & $ 0.12 (0.01)
 $ & $ 0.15 (0.02)
 $ & $ 0.17 (0.02)
 $ & $ 0.16 (0.01)
 $ & $ 0.19 (0.02)
 $ & $ 0.21 (0.02)
 $ & $ 0.20 (0.02)
 $ & $ 0.21 (0.02)
 $ & $ 0.20 (0.03)
 $ \\ \hline 
$ 0.0 $ & $ 16 $ & fc & $ 0.12 (0.01)
 $ & $ 0.16 (0.02)
 $ & $ 0.15 (0.02)
 $ & $ 0.22 (0.02)
 $ & $ 0.20 (0.02)
 $ & $ 0.22 (0.02)
 $ & $ 0.20 (0.02)
 $ & $ 0.23 (0.03)
 $ & $ 0.21 (0.04)
 $ \\ \hline \hline
$ 0.8 $ & $  8 $ & am & $ 0.268 (0.008)
 $ & $ 0.284 (0.007)
 $ & $ 0.281 (0.007)
 $ & $ 0.273 (0.006)
 $ & $ 0.251 (0.009)
 $ & $  $ & $  $ & $  $ & $  $ \\ \hline 
$ 0.8 $ & $  8 $ & fc & $ 0.281 (0.008)
 $ & $ 0.293 (0.007)
 $ & $ 0.281 (0.007)
 $ & $ 0.268 (0.006)
 $ & $ 0.26 (0.01)
 $ & $  $ & $  $ & $  $ & $  $ \\ \hline 
$ 0.8 $ & $ 12 $ & am & $ 0.26 (0.02)
 $ & $ 0.23 (0.02)
 $ & $ 0.25 (0.02)
 $ & $ 0.30 (0.02)
 $ & $ 0.28 (0.02)
 $ & $ 0.24 (0.01)
 $ & $ 0.28 (0.03)
 $ & $  $ & $  $ \\ \hline 
$ 0.8 $ & $ 12 $ & fc & $ 0.25 (0.02)
 $ & $ 0.27 (0.02)
 $ & $ 0.26 (0.02)
 $ & $ 0.30 (0.02)
 $ & $ 0.26 (0.02)
 $ & $ 0.25 (0.02)
 $ & $ 0.28 (0.02)
 $ & $  $ & $  $ \\ \hline 
$ 0.8 $ & $ 16 $ & am & $ 0.26 (0.04)
 $ & $ 0.25 (0.02)
 $ & $ 0.27 (0.02)
 $ & $ 0.39 (0.05)
 $ & $ 0.33 (0.04)
 $ & $ 0.25 (0.02)
 $ & $ 0.23 (0.03)
 $ & $ 0.26 (0.03)
 $ & $ 0.25 (0.04)
 $ \\ \hline 
$ 0.8 $ & $ 16 $ & fc & $ 0.23 (0.02)
 $ & $ 0.31 (0.04)
 $ & $ 0.30 (0.04)
 $ & $ 0.28 (0.04)
 $ & $ 0.26 (0.03)
 $ & $ 0.27 (0.04)
 $ & $ 0.25 (0.01)
 $ & $ 0.25 (0.03)
 $ & $ 0.28 (0.04)
 $ \\ \hline \hline
$ 1.6 $ & $  8 $ & am & $ 0.64 (0.03)
 $ & $ 0.61 (0.02)
 $ & $ 0.47 (0.02)
 $ & $ 0.31 (0.01)
 $ & $ 0.27 (0.01)
 $ & $  $ & $  $ & $  $ & $  $ \\ \hline
$ 1.6 $ & $  8 $ & fc & $ 0.67 (0.03)
 $ & $ 0.61 (0.02)
 $ & $ 0.47 (0.01)
 $ & $ 0.32 (0.01)
 $ & $ 0.28 (0.01)
 $ & $  $ & $  $ & $  $ & $  $ \\ \hline
$ 1.6 $ & $ 16 $ & am & $ 0.50 (0.04)
 $ & $ 0.55 (0.03)
 $ & $ 0.68 (0.04)
 $ & $ 0.49 (0.03)
 $ & $ 0.47 (0.03)
 $ & $ 0.33 (0.02)
 $ & $ 0.32 (0.02)
 $ & $ 0.29 (0.01)
 $ & $ 0.31 (0.03)
 $ \\ \hline 
$ 1.6 $ & $ 16 $ & fc & $ 0.59 (0.05)
 $ & $ 0.58 (0.03)
 $ & $ 0.57 (0.04)
 $ & $ 0.52 (0.05)
 $ & $ 0.45 (0.03)
 $ & $ 0.39 (0.03)
 $ & $ 0.31 (0.02)
 $ & $ 0.29 (0.02)
 $ & $ 0.29 (0.04)
 $ \\ \hline 
$ 1.6 $ & $ 24 $ & am & $ 0.57 (0.04)
 $ & $ 0.7 (0.1)
 $ & $ 0.60 (0.05)
 $ & $ 0.63 (0.04)
 $ & $ 0.50 (0.06)
 $ & $ 0.49 (0.05)
 $ & $ 0.48 (0.05)
 $ & $ 0.40 (0.09)
 $ & $ 0.35 (0.05)
 $ \\ \hline 
$ 1.6 $ & $ 24 $ & fc & $ 0.55 (0.04)
 $ & $ 0.60 (0.06)
 $ & $ 0.58 (0.09)
 $ & $ 0.58 (0.08)
 $ & $ 0.5 (0.1)
 $ & $ 0.42 (0.05)
 $ & $ 0.50 (0.09)
 $ & $ 0.53 (0.07)
 $ & $ 0.45 (0.08)
 $ \\ \hline  \hline
$ 2.7 $ & $  8 $ & am & $ 30.3 (0.5)
 $ & $ 1.9 (0.1)
 $ & $ 0.274 (0.005)
 $ & $ 0.141 (0.002)
 $ & $ 0.114 (0.003)
 $ & $  $ & $  $ & $  $ & $  $ \\ \hline 
$ 2.7 $ & $  8 $ & fc & $ 30.2 (0.5)
 $ & $ 2.0 (0.1)
 $ & $ 0.276 (0.005)
 $ & $ 0.142 (0.002)
 $ & $ 0.114 (0.003)
 $ & $  $ & $  $ & $  $ & $  $ \\ \hline 
$ 2.7 $ & $ 12 $ & am & $ 59 (2)
 $ & $ 4.2 (0.3)
 $ & $ 0.64 (0.02)
 $ & $ 0.275 (0.007)
 $ & $ 0.169 (0.004)
 $ & $ 0.132 (0.003)
 $ & $ 0.113 (0.004)
 $ & $  $ & $  $ \\ \hline 
$ 2.7 $ & $ 12 $ & fc & $ 58 (1)
 $ & $ 4.4 (0.3)
 $ & $ 0.64 (0.02)
 $ & $ 0.273 (0.007)
 $ & $ 0.171 (0.004)
 $ & $ 0.131 (0.003)
 $ & $ 0.114 (0.004)
 $ & $  $ & $  $ \\ \hline 
$ 2.7 $ & $ 16 $ & am & $ 94 (3)
 $ & $ 8.2 (0.8)
 $ & $ 1.19 (0.05)
 $ & $ 0.47 (0.01)
 $ & $ 0.253 (0.009)
 $ & $ 0.183 (0.006)
 $ & $ 0.144 (0.005)
 $ & $ 0.123 (0.004)
 $ & $ 0.116 (0.005)
 $ \\ \hline 
$ 2.7 $ & $ 16 $ & fc & $ 93 (3)
 $ & $ 8.2 (0.7)
 $ & $ 1.17 (0.04)
 $ & $ 0.47 (0.01)
 $ & $ 0.259 (0.009)
 $ & $ 0.183 (0.006)
 $ & $ 0.144 (0.005)
 $ & $ 0.123 (0.004)
 $ & $ 0.114 (0.005)
 $ \\ \hline \hline
\end{tabular}
\end{center}
\caption{~The four-momentum-space
          gluon propagator $D(k)$ [see eq.\
          \protect\reff{eq:D0def}
          and \protect\reff{eq:Dkdef}] as a function of
          $N k_{4}$.
          In all our runs we set $k = (0\mbox{,}\,
          0\mbox{,}\, 0\mbox{,}\, k_{4})$.
          Clearly the same value of $N k_{4}$ for different
          lattice sizes $N$ does not correspond to the same
          lattice momentum $p_{4}$ [see eq.\
          \protect\reff{eq:defp}].
          Since we use periodic boundary conditions,
          only data for $N k_{4} \leq N/2$
          are reported here.
          Two different types of statistics are considered:
          ``am'' and ``fc''. (For a
          definition see the last paragraph in Section
          \protect\ref{section:MLG}.) Error bars (in brackets) are
          one standard deviation.
          To make the table more readable,
          data for $N = 24$, $\beta = 1.6$ and $k_{4} > 8$
          are not reported.
}
\label{Table:gluon}
\vspace*{-0.3cm}
\end{table}

\clearpage

\clearpage

\clearpage
\begin{table}
\addtolength{\tabcolsep}{-1.0mm}
\hspace*{-1.0cm}
\protect\tiny 
\begin{center}
\begin{tabular}{|| c | c | c | c | c | c | c | c | c | c | c ||}
\hline
\hline
$ \beta $ & $ N $ & stat. &
$  1 $ & $  2 $ & $  3 $ & $  4 $ & $  5 $ & $  6 $ & $  7 $ & $  8 $
\\
\hline
\hline
$ 0.0 $ & $ 4 $ & am & $ 2.29 (0.03)
 $ & $ 0.894 (0.007)
 $ & $  $ & $  $ & $  $ & $  $ & $  $ & $  $ \\ \hline 
$ 0.0 $ & $ 4 $ & fc & $ 2.35 (0.03)
 $ & $ 0.916 (0.008)
 $ & $  $ & $  $ & $  $ & $  $ & $  $ & $  $ \\ \hline 
$ 0.0 $ & $ 6 $ & am & $ 5.21 (0.05)
 $ & $ 1.287 (0.009)
 $ & $ 0.876 (0.005)
 $ & $  $ & $  $ & $  $ & $  $ & $  $ \\ \hline 
$ 0.0 $ & $ 6 $ & fc & $ 5.54 (0.06)
 $ & $ 1.34 (0.01)
 $ & $ 0.906 (0.007)
 $ & $  $ & $  $ & $  $ & $  $ & $  $ \\ \hline 
$ 0.0 $ & $ 8 $ & am & $ 9.85 (0.08)
 $ & $ 2.19 (0.02)
 $ & $ 1.088 (0.005)
 $ & $ 0.880 (0.005)
 $ & $  $ & $  $ & $  $ & $  $ \\ \hline 
$ 0.0 $ & $ 8 $ & fc & $ 10.19 (0.08)
 $ & $ 2.23 (0.01)
 $ & $ 1.102 (0.004)
 $ & $ 0.889 (0.004)
 $ & $  $ & $  $ & $  $ & $  $ \\ \hline 
$ 0.0 $ & $ 10 $ & am & $ 16.0 (0.1)
 $ & $ 3.48 (0.02)
 $ & $ 1.534 (0.005)
 $ & $ 1.004 (0.003)
 $ & $ 0.877 (0.003)
 $ & $  $ & $  $ & $  $ \\ \hline 
$ 0.0 $ & $ 10 $ & fc & $ 16.9 (0.2)
 $ & $ 3.58 (0.02)
 $ & $ 1.564 (0.007)
 $ & $ 1.015 (0.004)
 $ & $ 0.888 (0.003)
 $ & $  $ & $  $ & $  $ \\ \hline 
$ 0.0 $ & $ 12 $ & am & $ 24.1 (0.2)
 $ & $ 5.19 (0.02)
 $ & $ 2.177 (0.008)
 $ & $ 1.285 (0.004)
 $ & $ 0.960 (0.002)
 $ & $ 0.876 (0.002)
 $ & $  $ & $  $ \\ \hline 
$ 0.0 $ & $ 12 $ & fc & $ 24.6 (0.2)
 $ & $ 5.28 (0.03)
 $ & $ 2.204 (0.009)
 $ & $ 1.296 (0.004)
 $ & $ 0.968 (0.003)
 $ & $ 0.883 (0.003)
 $ & $  $ & $  $ \\ \hline 
$ 0.0 $ & $ 14 $ & am & $ 33.1 (0.3)
 $ & $ 7.28 (0.04)
 $ & $ 2.99 (0.01)
 $ & $ 1.681 (0.007)
 $ & $ 1.157 (0.004)
 $ & $ 0.938 (0.003)
 $ & $ 0.876 (0.003)
 $ & $  $ \\ \hline 
$ 0.0 $ & $ 14 $ & fc & $ 34.8 (0.3)
 $ & $ 7.54 (0.04)
 $ & $ 3.07 (0.01)
 $ & $ 1.706 (0.006)
 $ & $ 1.170 (0.004)
 $ & $ 0.948 (0.003)
 $ & $ 0.888 (0.004)
 $ & $  $ \\ \hline 
$ 0.0 $ & $ 16 $ & am & $ 44.2 (0.7)
 $ & $ 9.9 (0.1)
 $ & $ 4.04 (0.04)
 $ & $ 2.20 (0.02)
 $ & $ 1.434 (0.006)
 $ & $ 1.085 (0.004)
 $ & $ 0.926 (0.004)
 $ & $ 0.879 (0.004)
 $ \\ \hline 
$ 0.0 $ & $ 16 $ & am & $ 45.5 (0.7)
 $ & $ 9.9 (0.1)
 $ & $ 4.02 (0.04)
 $ & $ 2.18 (0.02)
 $ & $ 1.432 (0.009)
 $ & $ 1.085 (0.006)
 $ & $ 0.922 (0.005)
 $ & $ 0.874 (0.005)
 $ \\ \hline  \hline
$ 0.8 $ & $  8 $ & am & $ 8.41 (0.07) $ & $ 1.778 (0.008) $ &
$ 0.861 (0.003) $ & $ 0.689 (0.002) $ & & & & \\
\hline
$ 0.8 $ & $  8 $ & fc & $ 8.94 (0.08) $ & $ 1.843 (0.009) $ &
$ 0.880 (0.003) $ & $ 0.701 (0.003) $ & & & & \\
\hline
$ 0.8 $ & $ 12 $ & am & $ 20.8 (0.1) $ & $ 4.42 (0.03) $ &
$ 1.793 (0.009) $ & $ 1.031 (0.004) $ & $ 0.761 (0.003) $ &
$ 0.690 (0.002) $ & &  \\
\hline
$ 0.8 $ & $ 12 $ & fc & $ 22.1 (0.6) $ & $ 4.53 (0.05) $ & $ 1.84 (0.01) $
& $ 1.053 (0.005) $ & $ 0.772 (0.003) $ & $ 0.697 (0.003) $ & & \\
\hline
$ 0.8 $ & $ 16 $ & am & $ 39.3 (0.6) $ & $ 8.47 (0.08) $ & $ 3.37 (0.02) $
& $ 1.800 (0.009) $ & $ 1.160 (0.006) $ & $ 0.866 (0.004) $ &
$ 0.732 (0.003) $ & $ 0.693 (0.003) $ \\
\hline
$ 0.8 $ & $ 16 $ & fc & $ 40.6 (0.7) $ & $ 8.7 (0.1) $ & $ 3.43 (0.03) $
& $ 1.82 (0.02) $ & $ 1.18 (0.01) $ & $ 0.875 (0.007) $ &
$ 0.738 (0.005) $ & $ 0.698 (0.004) $ \\
\hline
\hline
$ 1.6 $ & $ 8 $ & am &
$  6.61 (0.06) $ & $ 1.335 (0.006) $ & $ 0.632 (0.002) $ & $ 0.503 (0.001)
$ & & & & \\
\hline
$ 1.6 $ & $ 8 $ & fc &
$  7.2 (0.1) $ & $ 1.38 (0.01) $ & $ 0.645 (0.004) $ & $ 0.512 (0.003) $
& & & & \\
\hline
$ 1.6 $ & $ 16 $ & am &
$  32.2 (0.5) $ & $ 6.66 (0.07) $ & $ 2.56 (0.02) $ & $ 1.329 (0.005) $
& $ 0.845 (0.003) $ & $ 0.627 (0.002) $ & $ 0.528 (0.001) $ & $ 0.500 (0.001)
$ \\
\hline
$ 1.6 $ & $ 16 $ & fc &
$  32.1 (0.3) $ & $ 6.66 (0.04) $ & $ 2.56 (0.01) $ & $ 1.332 (0.006) $
& $ 0.847 (0.003) $ & $ 0.628 (0.002) $ & $ 0.530 (0.001) $ & $ 0.5004
(0.0009) $ \\
\hline
\hline
$ 2.7 $ & $ 8 $ & am &
$  3.4 (0.1) $ & $ 0.635 (0.002) $ & $ 0.3494 (0.0004) $ & $ 0.2937 (0.0003)
$ & & & & \\
\hline
$ 2.7 $ & $ 8 $ & fc &
$  3.4 (0.1) $ & $ 0.636 (0.002) $ & $ 0.3495 (0.0005) $ & $ 0.2938 (0.0003)
$ & & & & \\
\hline
$ 2.7 $ & $ 12 $ & am &
$  7.0 (0.2) $ & $ 1.308 (0.005) $ & $ 0.610 (0.001) $ & $ 0.3939 (0.0004) 
 $ & $ 0.3116 (0.0002) $ & $ 0.2893 (0.0002) $ & &  \\
\hline
$ 2.7 $ & $ 12 $ & fc &
$  7.1 (0.3) $ & $ 1.309 (0.005) $ & $ 0.610 (0.001) $ & $ 0.3939 (0.0004)
  $ & $ 0.3117 (0.0002) $ & $ 0.2894 (0.0002) $ & &  \\
\hline
$ 2.7 $ & $ 16 $ & am &
$  12.7 (0.5) $ & $ 2.31 (0.01) $ & $ 1.016 (0.002) $ & $ 0.6042 (0.0006)
  $ & $ 0.4272 (0.0003) $ & $ 0.3412 (0.0002) $ & $ 0.3004 (0.0001) $ &
$ 0.2883 (0.0001) $ \\
\hline
$ 2.7 $ & $ 16 $ & fc &
$  12.9 (0.6) $ & $ 2.32 (0.01) $ & $ 1.017 (0.002) $ & $ 0.6044 (0.0006)
  $ & $ 0.4273 (0.0003) $ & $ 0.3413 (0.0002) $ & $ 0.3004 (0.0001) $ &
$ 0.2883 (0.0001) $ \\
\hline
\hline
\end{tabular}
\end{center}
\caption{~The four-momentum-space ghost
          propagator $G(k)$ [see eq.\ \protect\reff{eq:Gk}]
          as a function of $N k_{4}$. 
          In all our runs we set $k = (0\mbox{,}\,
          0\mbox{,}\, 0\mbox{,}\, k_{4})$.
          Clearly the same value of $N k_{4}$ for different
          lattice sizes $N$ does not correspond to the same
          lattice momentum $p_{4}$ [see eq.\
          \protect\reff{eq:defp}].
          Since we use periodic boundary conditions,
          only data for $N k_{4} \leq N/2$
          are reported here.
          Two different types of statistics are considered:
          ``am'' and ``fc''. (For a definition see the last paragraph in 
          Section
          \protect\ref{section:MLG}.) Error bars (in brackets) are
          one standard deviation.
}
\label{Table:ghost}
\vspace*{0.3cm}
\end{table}

\clearpage

\clearpage

\begin{table}
%
\hspace*{-1.0cm}
\protect\footnotesize
\begin{center}
\begin{tabular}{|| c | c | c ||}
\hline
\hline
$ \beta $ & $ N $ & $s$
\\
\hline
$ 0.0 $ & $ 4 $ & $ 1.35 (0.03) $ \\
\hline
$ 0.0 $ & $ 6 $ & $ 1.27 (0.01) $ \\
\hline
$ 0.0 $ & $ 8 $ & $ 1.23 (0.01) $ \\
\hline
$ 0.0 $ & $ 10 $ & $ 1.187 (0.009) $ \\
\hline
$ 0.0 $ & $ 12 $ & $ 1.166 (0.009) $ \\
\hline
$ 0.0 $ & $ 14 $ & $ 1.135 (0.001) $ \\
\hline
$ 0.0 $ & $ 16 $ & $ 1.11 (0.02) $ \\
\hline
\hline
$ 0.8 $ & $  8 $ & $ 1.27 (0.01) $
\\
\hline
$ 0.8 $ & $ 12 $ & $ 1.18 (0.01) $
\\
\hline
$ 0.8 $ & $ 16 $ & $ 1.13 (0.02) $
\\
\hline
\hline
$ 1.6 $ & $  8 $ & $ 1.30 (0.01) $
\\
\hline
$ 1.6 $ & $ 16 $ & $ 1.17 (0.02) $
\\
\hline
\hline
$ 2.7 $ & $  8 $ & $ 1.36 (0.03) $
\\
\hline
$ 2.7 $ & $ 12 $ & $ 1.27 (0.03) $
\\
\hline
$ 2.7 $ & $ 16 $ & $ 1.27 (0.03) $
\\
\hline
\hline
\end{tabular}
\end{center}
\caption{~Estimate of the pole of the ghost propagator $G(k)$
           [see eq.\ \protect\reff{eq:Gk}] as the momentum
           goes to zero. The values of the exponent $s$ [see eq.\
           \protect\reff{eq:defs}] are obtained
           from the data corresponding
           to the two smallest momenta available
           for each lattice size
           (Table \protect\ref{Table:ghost}, average ``am'').
           Errors bars (in brackets)
           come from propagation of the statistical error on 
           ghost propagator.
}
\label{Table:s2}
\vspace*{0.3cm}
\end{table}

\end{document}